# Generalization of differential Maxwell equations from integral electrodynamics laws for moving ambiences.

V.Bashkov, S.Kozyrev
*Department of physic, Kazan State University*
e-mail: Sergeym@space.com, Victor.Bashkov@ksu.ru.

In this work we obtained differential equations, describing interaction of electromagnetic field with moving sidebars and surfaces, coming from integral electrodynamics laws. Herewith, it has been shown that received differential equations contain but the such features of electromagnetic field as **E, H,** distributions of charges and currents, as well as features of surfaces, to which sidebar is leaning, changing its form with a time. These features are field of normals to surfaces **N** and field of velocities of changing these normals with a time d**N**/dt.

We assume, that main laws of electromagnetic field are written in the integral form as follows [1]:

1. **The First electrodynamics law**. Flow of electrical field of the tension **E** through any closed surface is equal to charge, disposed inside of the surface:

$$\oiint_{\Sigma} (\vec{B}\,\vec{n})\, d\sigma = \int\int\int_{V} \frac{d\,q_V}{\epsilon_0} \tag{1}$$

2. **The Second electrodynamics law**. For any closed curve in three-dimensional space a circulation of electrical field is equal to velocities of changing a flow of magnetic field **H** through the surface strained on this sidebar.

$$\oint_{L} \vec{E}\, d\,\vec{r} = \frac{d}{dt}\Big[\int\int_{\Sigma} (\vec{B}\,\vec{n})\, d\sigma\Big] \tag{2}$$

3. **The Third electrodynamics law**. Flow of the tension of magnetic field **H** through any closed surface is a zero.

$$\oiint_{\Sigma} (\vec{B}\,\vec{n})\, d\sigma = 0 \tag{3}$$

4. **The Fourth electrodynamics law**. For any surface Σ, limited by the closed sidebar L, circulation of vector **B** along any closed regular curve equals to the velocity of changing of flow of the tension of electrical field **E** through any surface, leaning on this curve plus electrical current, run through this surface.

$$c^2 \oint_{L} \vec{B}\, d\,\vec{r} = \frac{d}{dt} \int\int_{\Sigma} (\vec{E}\,\vec{n})\, d\sigma + \frac{d}{dt}\Big[\int\int_{\Sigma} dq_{\Sigma}\Big] \tag{4}$$

Here c - velocity of light and

$$dq_{\Sigma} = \rho\, d\sigma \tag{5}$$

where ρ density of charges on Σ.

We have written last expression in the correlation (4) in the manner different from considered in [1] and hence demands to fix a charged surface, with the mass surface in analogy introduced in monographs [2].

Then

$$\frac{d}{dt}\Big[\int\int_{\Sigma} dq_{\Sigma}\Big] = \frac{d}{dt}\Big[\int\int \rho\, d\sigma\Big] \tag{6}$$

where

$$d\sigma = \left|\vec{N}\right| du\, dv \tag{7}$$

$\vec{N}$ is vector of normal to surface Σ generally dependent on time t.

Using a Gauss law brings for equations (1) and (3) to the following differential laws of electromagnetism:

$$\operatorname{div} \vec{E} = 4\pi\rho \tag{1'}$$

$$\operatorname{div} \vec{B} = 0 \tag{2'}$$

However, direct using of Stocks and Gauss formulae for cases, when type of surface, the velocity of changing of flow to be find through, depends on time here are not applicable, since it is impossible to change places of operation of integrating on surfaces and differentiation on a time.

Let's make first formally operation

$$\frac{d}{dt}\left[\iint_{\Sigma(t)} \left(\vec{A}\,\vec{n}\right) d\sigma\right] = \frac{d}{dt}\left[\iint_{\Omega} \left(\vec{A}\,\vec{N}\right) du\,dv\right] \tag{9}$$

in the case of fixed surface Σ in the parametric type (8). **A** - is a certain vector field. Then, in (9)

$$\vec{N} = \vec{r}_u \times \vec{r}_v = \begin{vmatrix} \vec{i} & \vec{j} & \vec{k} \\ \frac{\partial x}{\partial u} & \frac{\partial y}{\partial u} & \frac{\partial z}{\partial u} \\ \frac{\partial x}{\partial v} & \frac{\partial y}{\partial v} & \frac{\partial z}{\partial v} \end{vmatrix} \tag{10}$$

$$u_0 \leq u \leq u_1,\ v_0 \leq v \leq v_1 \tag{11}$$

Moreover **u** and **v** already are already independent on time. Note that if we use a dependency from time of surface Σ in the manner of more complex than (8) then it brings to difficulties, for details that we refer a reader to monographs [2], where there is an extensive bibliography on this subject.
Further we get

$$\frac{d}{dt}\iint_{\Omega} \left(\vec{A}\,\vec{N}\right) du\,dv = \iint_{\Omega} \frac{d}{dt}\left(\vec{A}\,\vec{N}\right) du\,dv \tag{11}$$

Let's calculate

$$\frac{d}{dt}\left(\vec{A}\,\vec{N}\right) = \lim_{\Delta t \to 0}\left[\frac{\left(\vec{A}\,\vec{N}\right)\begin{matrix} x\,(t+\Delta t,\ u,\ v) \\ y\,(t+\Delta t,\ u,\ v) \\ z\,(t+\Delta t,\ u,\ v)\end{matrix} - \left(\vec{A}\,\vec{N}\right)\begin{matrix} x\,(t,\ u,\ v) \\ y\,(t,\ u,\ v) \\ z\,(t,\ u,\ v)\end{matrix}}{\Delta t}\right] = \tag{12}$$

$$\frac{\partial\left(\vec{A}\,\vec{N}\right)}{\partial t} + \left(\vec{V}\,\vec{\nabla}\right)\left(\vec{A}\,\vec{N}\right) = \left(\dot{\vec{A}}\,\vec{N} + \vec{A}\,\dot{\vec{N}}\right)$$

where

$$\vec{V} = \frac{\partial\,\vec{r}\,(t,\ u,\ v)}{\partial\,t}$$

is velocity of moving of points on surface. Lowering intermediate calculations we will find

$$\frac{d}{dt}\left[\iint_{\Sigma(t)} \left(\vec{A}\,\vec{n}\right) d\sigma\right] = \frac{d}{dt}\left[\iint_{\Omega} \left(\vec{A}\,\vec{N}\right) du\,dv\right] = \iint_{\Omega} \left(\dot{\vec{A}}\,\vec{N} + \vec{A}\,\dot{\vec{N}}\right) du\,dv \tag{13}$$

Similarly

$$\begin{aligned} \frac{d}{dt}\left(\rho\,\left|\vec{N}\right|\right) &= \frac{\partial}{\partial\,t}\left(\rho\,\left|\vec{N}\right|\right) + \left(\vec{V}\,\vec{\nabla}\left(\rho\,\left|\vec{N}\right|\right)\right) = \\ &\frac{\partial\,\rho}{\partial\,t}\left|\vec{N}\right| + \rho\,\frac{\partial\,\left|\vec{N}\right|}{\partial\,t} + \left(\left(\vec{V}\,\vec{\nabla}\right)\rho\right)\left|\vec{N}\right| + \rho\,\left(\left(\vec{V}\,\vec{\nabla}\right)\left|\vec{N}\right|\right) \end{aligned} \tag{14}$$

Applying Stocks and Gauss theorems to (2) and (4) and taking into account (13) and (14) we get:

$$\left(\text{rot}\,\vec{E} \cdot \vec{N}\right)_t = \left(\dot{\vec{B}}\,\vec{N} + \vec{B}\,\dot{\vec{N}}\right) \tag{2'}$$

$$c^2\left(\text{rot}\,\vec{B} \cdot \vec{N}\right) = \left(\dot{\vec{E}}\,\vec{N} + \vec{E}\,\dot{\vec{N}}\right) + \dot{\rho}\left|\vec{N}\right| + \rho\left|\dot{\vec{N}}\right| \tag{4'}$$

Let's consider an application of received equations (1') - (4') for the partial case [1], [3]. On the Pic.1 it is shown a simple wire loop, which sizes are to be change with a time.

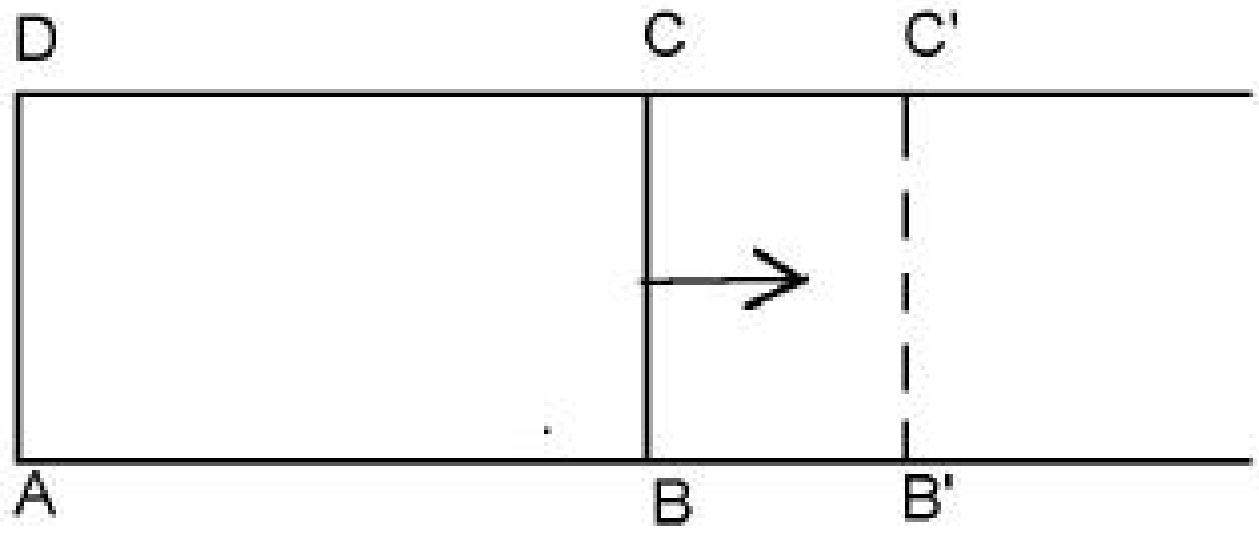

**Pic.1**

Equation of plane strained on loop A B' C' D is

$$\sum\nolimits_t \begin{cases} x = ut \\ y = v \\ z = 0 \end{cases} \qquad (15)$$

where $\mathbf{0 \le u \le u_1}$, $\mathbf{0 \le v \le v_1}$ or $\mathbf{v} = \dot{\vec{r}}(t, u, v) = u\,t\,\mathbf{i} + v\,\mathbf{j}$

Equation of loop A B' C' D = A B' ∪ B' C ∪ C' D ∪ D A is:

$$A B' : \begin{matrix} x = ut;\ 0 \le u \le u_1 \\ y = 0 \end{matrix} ;\quad B' C' : \begin{matrix} x = u_1 t \\ 0 \le v \le v_1 \end{matrix} \qquad (16)$$

$$C' D' : \begin{matrix} x = ut;\ 0 \le u \le u_1 \\ y = v_1 \end{matrix} ;\quad D A : \begin{matrix} x = 0 \\ y = v;\ 0 \le v \le v_1 \end{matrix}$$

Equations of electromagnetic field

$$\oint_L \vec{E}\, d\vec{r} = \frac{d}{dt} \iint_{\Sigma_t} \left(\vec{B} \cdot \vec{n}\right) d\sigma = \iint_{\Omega} \left(\vec{B} \cdot \vec{N}\right)^{\bullet}_t du\, dv \qquad (17)$$

for our case bring to:

$$\int_0^{u_1} \overset{(1)}{E}\, t\, du + \int_0^{v_1} \overset{(2)}{E}\, dv + \int_{u_1}^{0} \overset{(3)}{E}\, t\, du + \int_{v_1}^{0} \overset{(4)}{E}\, dv = \qquad (18)$$

$$\iint_{\substack{0 \le u \le u_1 \\ 0 \le v \le v_1}} B_0\, du\, dv = B_0\, u_1 v_1 = B_0 S$$

or

$$\overset{(1)}{E}\, l_1 + \overset{(2)}{E}\, l_2 + \overset{(3)}{E}\, l_3 + \overset{(4)}{E}\, l_4 = B_0 S \qquad (19)$$

where $S = u_1 v_1$

As far as loop consists of conducting wire, when changing an area covering a loop under the action of magnetic field $B_0$ a current appears

$$\vec{j} = \vec{j}_1 = \vec{j}_2 = \vec{j}_3 = \vec{j}_4 \qquad (20)$$

being

$$|\, j_i \,|_{i=1,2,3,4} = \frac{v_i}{R_i} = \frac{\left|\overset{(i)}{E}\right| l_i}{R_i} = \frac{1}{\rho_0} \left|\overset{(i)}{E}\right| \qquad (21)$$

where $\rho_0$ – is a resistance of the unit of length. From (20), (21) we get, that

$$\left|\overset{(1)}{E}\right| = \left|\overset{(2)}{E}\right| = \left|\overset{(3)}{E}\right| = \left|\overset{(4)}{E}\right|$$

and hence

$$E = \left| \overset{(i)}{E} \right| = \frac{B_0 S}{L} = \frac{B_0 u_1 v_1}{u_1 t + v_1}$$

It is easy show that the result does not depend on forms of surface strained on the loop.

Obviously, when **V** = $\dot{\mathbf{N}}$ = 0, i.e. in the case of immobile surface Σ the equations (1'), (2'), (3') and (4') are reduced to Maxwell equations for the electromagnetic field [1].

$$\mathrm{div}\,\vec{E} = \frac{\rho}{\epsilon_0} \qquad (1'')$$

$$\mathrm{div}\,\vec{H} = 0 \qquad (2'')$$

$$\mathrm{rot}\,\vec{E} = \frac{\partial \vec{B}}{\partial t} \qquad (3'')$$

$$c^2\,\mathrm{rot}\,\vec{H} = \frac{\partial \vec{E}}{\partial t} + \frac{\vec{j}}{\epsilon_0} \qquad (4'')$$

due to the **N** is arbitrary**.**

Considering this example it was important to fix a surface, covering by the loop in the manner of (1), which shows that this piece of surface is being sprained or compressed and hence produce expansion or compression of ambience (ethers), covered by the loop. Length of loop with coordinates **x = 0, y = v**, **0 £ v £ v$_1$** did not change its position and at the same time other lengths of loop, including border length **x = v$_1$, y = v**, **0 £ v £ v$_1$** were moving. Influence of constant magnetic field exactly on this expanding and compressing ambience, covered by the loop, and brings to arising of electrical field in each point of surface, covering by the loop, and arising of current in the loop.

If we would have assigned onward moving of a loop as whole in space given by equation

**x = u +w t** **0 £ u £ u$_1$**

**y = v** **0 £ v £ v$_1$**

then magnetic field did not create on surfaces covering by the loop no electrical field and consequently current in the loop. This fact is confirmed on experiment [1], [3]. So, our generalization of differential Maxwell equations on the case of expansion or compression of ethers - a carrier of electromagnetic field will allow to get additional information on characteristics of this field and its carrier - an ether.

In conclusion it's possible to notice the following: the velocity of moving of ambience **V** is not connected with the velocity of spreading of electromagnetic field with in vacuum. If ambience is a carrier of electromagnetic field is move, the question of invariance of received equations will be a subject of the following article. There we will show that vector **N** normal to the moving surface can changed with velocity greater then velocity of light.

Ideas of present work are published earlier in [4].